\documentclass[a4paper,fleqn,usenatbib]{mnras}

\usepackage{footnote}
\usepackage[T1]{fontenc}
\usepackage{ae,aecompl}
\usepackage[flushleft]{threeparttable}
\usepackage{longtable,threeparttablex,booktabs,threeparttable}
\usepackage{longtable}
\usepackage[flushleft]{threeparttable}
\usepackage{longtable,threeparttablex,booktabs,threeparttable}
\usepackage{graphicx}
\usepackage{wrapfig}
\usepackage{pdflscape}
\usepackage{rotating}
\usepackage{epstopdf}
\usepackage{supertabular}

\usepackage{graphicx}	
\usepackage{amsmath}	
\usepackage{amssymb}	
\usepackage{multirow,multicol}
\newcommand{\HI}{H{\sc i}}

\newcommand{\hi}{H{\sc i}~21-cm }

\title[\hi absorption in intermediate redshift galaxies]{uGMRT detections of \hi absorption associated with intermediate redshift galaxies}

\author[Aditya]{J. N. H. S. Aditya$^{1}$\thanks{adityaj@iucaa.in} \\
$^{1}$Inter-University Centre for Astronomy and Astrophysics, Pune 411007, India\\
\newline
}

\date{Accepted XXX. Received YYY; in original form ZZZ}

\pubyear{2018}

\begin{document}
\label{firstpage}
\pagerange{\pageref{firstpage}--\pageref{lastpage}}
\maketitle

\begin{abstract}

I report detections of four new \hi absorbers associated with sources at intermediate redshifts, $0.7 < z < 1.0$. The sources are part of a sample of 11 radio-loud galaxies, all at $0.7 < z < 1.0$, 
that were searched for associated \hi absorption using uGMRT.
Previously, just four such absorbers were known in the literature at these redshifts; the current observations have 
increased the total to eight. The results indicate that the detection fraction at intermediate redshifts could be as high as that at lower redshifts, $\approx 30\%$, on contrary to a much lower detection fraction observed in samples at $z > 1$. Three detections show strong blueshifted features, indicating cold gas outflows. These three sources also tentatively show excess [O {\sc ii}] line luminosity compared to a bulk of the remaining sample, possibly suggesting that the hosts of these AGNs harbour different environments, either due to interaction with the radio jets or due to excess star formation in the host galaxy. Further, a cold \HI~mass outflow rate of $\approx 78$~{\rm M}$_{\odot}$~yr$^{-1}$, assuming $\rm {T_{s} = 1000}$~K, is estimated for the detection towards SDSS J014652.79-015721.2, at $z = 0.95904$, which is the highest till date in comparison to similar estimates available in the literature.

\end{abstract}



\section{INTRODUCTION}

The evolutions of the Active Galactic Nucleus (AGN) and its host galaxy are thought to be closely related \citep[e.g.,][]{ferrarese2000, wang2013}. The fuelling of AGN activity and the feedback mechanisms are two important processes that likely involve significant interactions between the AGN and its host galaxy, and have strong infuence on galaxy evolution. The circumnuclear material that constitutes ionized, atomic and molecular gas, and dust components, is thought to provide the fuel for the AGN activity \citep[e.g.,][]{heckman1986, vangorkom89, canalizo2007}. The gas inflows to the central regions either through slow accretion, or through triggered infalls due to galaxy mergers that often lead to circumnuclear starbursts \citep[e.g.,][]{jogee2006, davies2010}. Conversely, there has been increasing evidence in recent years showing that the AGN jets kinematically interact with the surrounding medium, creating outflows of cold and/or hot gas \citep[e.g.,][]{morganti2013, emonts2014, tombesi2014}. Such mechanisms could lead to quenching of star formation in the host galaxy and also could end the active state of the nucleus in a few cases \citep[e.g.,][]{fabian2012}. Discerning the distribution and kinematical properties of the gaseous material in the vicinities of AGNs is hence essential for understanding galaxy evolution.  

Atomic hydrogen (\HI), a dominant component of the gaseous interstellar medium, can be probed via \hi absorption against the bright central radio source, the AGN. A majority of such `associated' \hi absorption studies in the past are mainly focused
at low redshifts, $z < 1$. For example, \citet[][]{vermeulen2003} conducted a survey of compact objects in the redshift range $0.2 < z < 0.84$. Later, \citet[][]{gupta2006} have searched for associated \hi absorption towards a sample of AGNs, in which a majority of the sources were at redshifts $z < 0.35$. Recently, \citet[][]{gereb2015} and \citet[][]{maccagni2017} observed a large sample of $\sim 250$ sources at redshifts $z < 0.3$. The kinematics of the gas relative to the AGN can be inferred through the \hi absorption technique. The above studies find that narrow absorption profiles (with widths $\lesssim 200~ \rm{km~s^{-1}}$) that are close to the AGN systemic velocity, are mostly produced by \HI~clouds rotating in circumnuclear disks, while broader profiles with blueshifted components indicate unsettled gas structures that are possibly the resultants of gas-rich mergers or strong outflows. Particularly, wide (with width $\rm \gtrsim 300~km~s^{-1}$) and shallow blue-shifted wings are often found to be associated with jet driven outflows, because VLBI follow-up observations could trace these wings against the radio jets, and associate the high velocity gas with the expansion of the AGN \citep[e.g.][]{oosterloo2000, morganti2013}. Here, the redshift of the AGN is most often estimated through optical emission lines.

Overall, at present over 650 systems have been searched for associated \hi absorption \citep[e.g.,][]{vermeulen2003, gupta2006, curran2013, aditya2016, maccagni2017, aditya2018}, with more than 120 detections reported in the literature. A majority ( $\gtrsim 550$) of the searches have been limited to redshifts $z < 0.6$, that consist of samples of 
both compact, flat-to-inverted spectrum sources and extended,
steep-spectrum radio sources. The typically reported detection fraction at these low redshifts is 
$\approx 30 \%$. At redshifts $z > 1.0$, however, compact, flat-spectrum radio sources dominate the samples
that have been searched for associated \hi absorption. Approximately 90 sources have been searched at $z > 1.0$ 
\citep[e.g.,][]{curran2013, aditya2016, aditya2018} with the detection rate being consistently low at $\lesssim 10\%$ 
(compared to $\sim 30\%$ at low-$z$ - cf. \citealp[][]{gereb2015} ). Further, \citet[][]{aditya2016, aditya2018} reported that among flat-spectrum radio sources the strength of \hi absorption is weaker at higher redshifts. The difference in the
absorption strengths in the low-$z$ and high-$z$ sub-samples was found to be of $3\sigma$ significance. 
Although the low detection rate at $z > 1$ indicates a 
possible redshift evolution in the gaseous environments of AGNs, it is important to note that the high-$z$ samples are 
currently dominated by flat-spectrum radio sources whereas low-$z$ samples include both flat and steep spectrum sources.
It hence needs to be tested whether the results pertain to all classes of AGNs.

Presently, the intermediate redshift range, i.e. $0.7 < z < 1.0$, has just 14 searches for associated \hi absorption reported in the literature \citep[][]{carilli1998, peck2000, vermeulen2003, curran2011a, salter2010, yan2016}. Moreover, the number of reported detections is just four. The physical conditions of neutral gas in systems at these intermediate redshifts are largely unprobed. The lack of suitable observing bands
in the currently available telescopes is the main hindrance for \hi observations in this range. With the advent of uGMRT (upgraded Giant Metre Wave Radio Telescope), that covers almost seamlessly from 50 MHz to 1500 MHz, it is now possible to  study \hi absorption systems up to very high redshifts.
The newly released 550-850 MHz (Band-4) receiver is particularly suitable for \hi absorption studies at intermediate to high redshifts, $0.7 < z < 1.6$.
Here, I report observations from a pilot survey to search for \hi absorption in a sample of 11 radio galaxies at $0.7 < z < 1$, using recently released Band-4 of the uGMRT.

\begin{table*}
\footnotesize
\caption{The sample of 11 radio galaxies.} 
\label{17sam}
\begin{scriptsize}
\begin{tabular}{lllllllll}
\hline 
Source &  Alt.    & $z$     & S (FIRST)  & S (FIRST)       &  W1    &  W2    &  W3     &  W4    \\
 name  &  name    &         &  peak      &  integrated     &        &        &         &        \\
       &          &         & [mJy]      &  [mJy]          & [mag]  & [mag]  & [mag]   & [mag]  \\
\hline

 SDSS J014652.79-015721.2    & 0146-01    & 0.95904     &   861.2   &  882.5  &   16.54   &    15.99   &   12.39 & 8.41     \\
 SDSS J075409.26+411000.0    & 0754+41    & 0.87017     &   203.9   &  208.8  &   15.33   &    14.05   &   10.88 & 8.30     \\
 SDSS J091022.55+241919.5    & 0910+24    & 0.90669     &   568.0   &  827.5  &   15.78   &    15.17   &   11.21 & 8.56     \\
 SDSS J091022.86+001935.4    & 0910+00    & 0.95937$^a$ &   288.6   &  333.9  &   15.64   &    15.22   &   12.33 & 9.07     \\
 SDSS J100631.76+171317.0    & 1006+17    & 0.82148     &   567.3   &  584.6  &   14.51   &    13.59   &   10.99 & 8.98     \\
 SDSS J101301.60+244837.3    & 1013+24    & 0.94959     &   511.4   &  541.7  &   15.11   &    14.34   &   11.64 & 8.87     \\
 SDSS J104830.37+353800.8    & 1048+35    & 0.84644     &   269.2   &  309.3  &   14.37   &    13.30   &   10.45 & 8.23     \\
 SDSS J130407.32+370908.1    & 1304+37    & 0.93987$^a$ &   284.9   &  305.8  &   13.72   &    12.73   &    9.96 & 7.72     \\
 SDSS J144850.36+040219.9    & 1448+04    & 0.87107     &   352.9   &  377.5  &   14.88   &    14.06   &   11.38 & 8.43     \\
 SDSS J150619.62+093451.7    & 1506+09    & 0.80772     &   677.2   &  691.7  &   15.15   &    14.89   &   11.80 & 8.39     \\
 SDSS J164801.53+222433.3    & 1648+22    & 0.82266     &   250.4   &  253.0  &   15.27   &    14.54   &   11.49 & 8.63     \\

\hline
\hline
\end{tabular}
\end{scriptsize}
\begin{tablenotes}

\item[]~Notes:  The parameters in the table are (1) the source name, (2) the alternative name, for easy recognition of sources in this paper, (3) the redshift from SDSS DR14, (4) and (5) the peak and integrated flux densities at 1.4~GHz, in mJy, obtained from FIRST catalogue, (6), (7), (8) and (9) the magnitudes at 3.4 $\mu$m, 4.6 $\mu$m, 12 $\mu$m and 22 $\mu$m bands, respectively, obtained from WISE infrared catalogue. 
\newline
$^a$ Note: For these two sources, the emission lines detected in the SDSS spectra are weak leading to relatively larger redshift uncertainties.
 
\end{tablenotes}
\end{table*}

 \hspace{-1cm}
\begin{table*}
\footnotesize
\caption{Observational details and results of 11 sources.} 
\label{17gal}
\begin{scriptsize}
\begin{tabular}{lllllllllllll}
\hline \\
Source &  $z$ & $\nu_{HI}$ &   Spectral      & S$_{obs}$  & $\Delta $S  &  $\Delta \tau$    & $\int \tau dv$$^a$  & $\rm{ N_{HI} }$$^a$  & ${\rm L}'_{1.4~GHz}$$^{b}$ & $\rm{ L_{OII}}$ \\
       &      &            &   resolution    &            &             &  $\times 10^{-3}$ &                     & $\rm \times 10^{18}$  &                           & $\times 10^{41}$\\
       &      &            &                 &            &             &                   &                     & $\rm \times(T_{s} /cf)$&                          &                 \\
       &      & [MHz]      &   [km s$^{-1}$] & [mJy]      & [mJy]       &                   & [km s$^{-1}$]       & [cm$^{-2}$]          &                            & [erg s$^{-1}$] \\
\hline

0146-01$^c$   &  0.95904 &  725.05  &  13.5$^d$  &  $1803.8$   &  2.59  &  1.44  &  $11.46  \pm  0.22$  & $20.89 \pm 0.04$ &       27.65                 & $185.7 \pm 1.9$ \\
0754+41       &  0.87017 &  759.51  &  19.3  &  $368.8$    &  4.12  &  11.18 &  $<3.35$             & $<6.11$          &       26.88                 & $28.1 \pm 0.6$  \\
0910+24       &  0.90669 &  744.96  &  19.6  &  $1229.3$   &  6.35  &  5.20  &  $<1.55$             & $< 2.83$         &       27.43                 & $163.7 \pm 0.9$ \\
0910+00       &  0.95937 &  724.93  &  20.2  &  $ 504.9 $  &  3.43  &  6.80  &  $< 2.04$            & $<3.72$          &       27.09                 & $12.7 \pm 0.6$  \\
1006+17       &  0.82148 &  779.81  &  18.8  &  $509.9 $   &  5.47  &  10.7  &  $<3.22$             & $<5.87$          &       26.96                 & $14.3 \pm 0.6$  \\
1013+24$^c$   &  0.94959 &  728.57  &  3.4$^d$   &  $891.9 $   &  1.23  &  1.37  &  $1.15 \pm 0.05$     & $2.10 \pm 0.09$  &       27.33                 & $33.1 \pm 0.9$  \\
1048+35$^c$   &  0.84644 &  769.27  &  3.2$^d$   &  $552.9 $   &  1.82  &  3.28  &  $4.77 \pm 0.18$     & $8.69 \pm 0.33$  &       27.03                 & $100.2 \pm 1.1$ \\
1304+37       &  0.93987 &  732.22  &  19.9  &  $484.4 $   &  2.28  &  4.71  &  $<1.41$             & $< 2.57$         &       27.06                 & $17.7 \pm 1.2$  \\
1448+04       &  0.87107 &  759.14  &  19.3  &  $476.2 $   &  3.81  &  8.0   &  $<2.40$             & $<4.38$          &       26.99                 & $30.0 \pm 0.9$  \\
1506+09       &  0.80772 &  785.74  &  18.6  &  $1289.1 $  &  4.48  &  3.48  &  $<1.04$             & $<1.89$          &       27.35                 & $8.8 \pm 0.4$   \\
1648+22       &  0.82266 &  779.30  &  9.4$^d$   &  $252.8 $   &  3.37  &  13.15 &  $16.06 \pm 0.74$    & $29.28 \pm 1.35$ &       26.66                 & $3.9 \pm 0.3$   \\

\hline
\hline
\end{tabular}
\end{scriptsize}
\begin{tablenotes}

\item[]~Notes:  The parameters in the table are (1) the source name, (2) the source redshift, $z$, (3) the redshifted \hi line frequency, $\nu_{HI}$, in MHz, (4) the spectral resolution, in km s$^{-1}$, after Hanning-smoothing and re-sampling the spectra, (5) the observed flux density of the source, S$_{obs}$, in mJy, (6) the RMS noise on the spectrum for non-detections, and on the line-free channels for spectra with detections, $\Delta $S, in mJy, (7) the optical depth RMS on the spectrum for non-detections, and on the line-free channels for spectra with detections, $\Delta \tau$, (8) the velocity integrated optical depth of the \hi absorption, $\int \tau dv$, in km s$^{-1}$, (9) the \HI~column density, $\rm{ N_{HI} }$, (10) the logarithm of the AGN luminosity
$\rm{L_{1.4~GHz}}$ in W Hz$^{-1}$ at a rest-frame frequency of 1.4 GHz, i.e. ${\rm L}'_{1.4~GHz}$ = Log[${\rm L}_{1.4~GHz}$ / W Hz$^{-1}$], and
(11) the [O {\sc ii}] line luminosity, $\rm{ L_{OII}}$, in erg s$^{-1}$.
\newline
$^a$ Note: For \hi non-detections, the values are 3$\sigma$ upper limits on the velocity-integrated optical depth and the \HI~column density, assuming a Gaussian line with a FWHM of 100 km s$^{-1}$. Here, $\rm T_{s}$ is the spin temperature and $\rm cf$ is the covering factor. 
\newline
$^b$ Note: ${\rm L}'_{1.4~GHz}$ = Log[${\rm L}_{1.4~GHz}$ / W Hz$^{-1}$].
\newline
$^c$ Note: For these sources, the details correspond to the second observing run, that has a higher sensitivity. 
\newline
$^d$ Note: For these sources, the quoted velocity resolution is without Hanning-smoothing and re-sampling.
\end{tablenotes}
\end{table*}

\begin{figure}

\includegraphics[height=10cm, width=9cm]{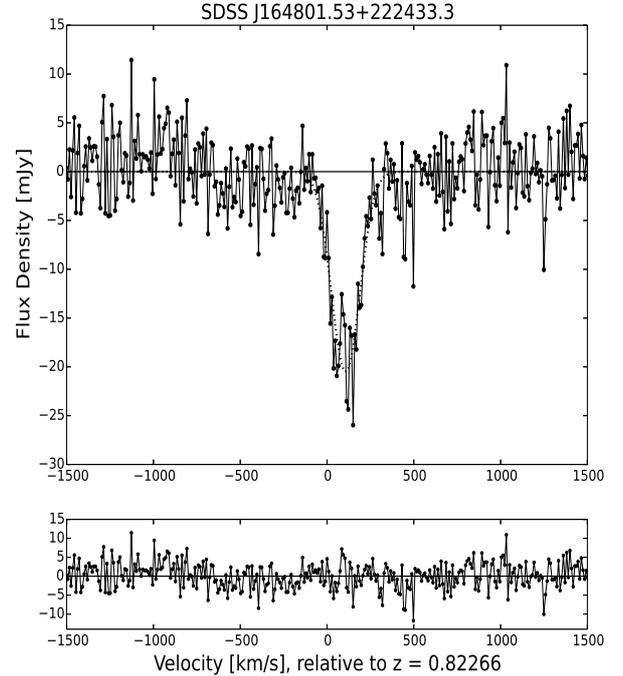} \\
\caption[]{The uGMRT detection of associated \hi absorption towards 1648+22, at $z = 0.82266$. Top panel shows the single Gaussian fit 
to the absorption profile, and the panel at bottom shows the residuals after the fit has been subtracted from the spectrum.}\label{det1}
\end{figure}

\begin{figure}
\begin{tabular}{c}
\includegraphics[height=10cm, width=9cm]{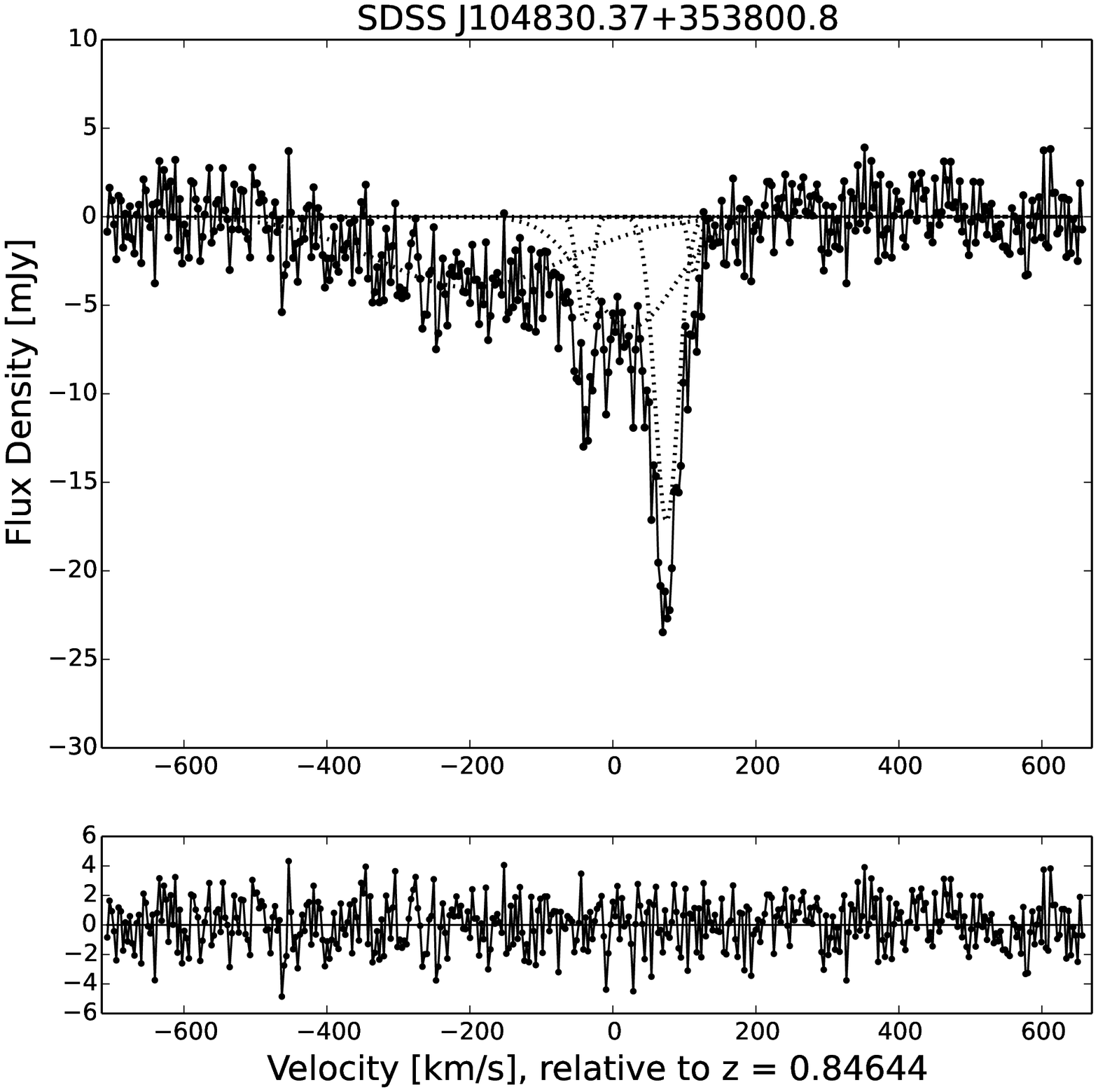} \\ 
\end{tabular}
\caption[]{The uGMRT detection of associated \hi absorption towards 1048+35, at $z = 0.84644$. Top panel shows the Gaussian fits 
to the absorption profile, and the panel at bottom shows the residuals after the fit has been subtracted from the spectrum. 
}\label{det2}
\end{figure}

\begin{figure}

\includegraphics[height=10cm, width=9cm]{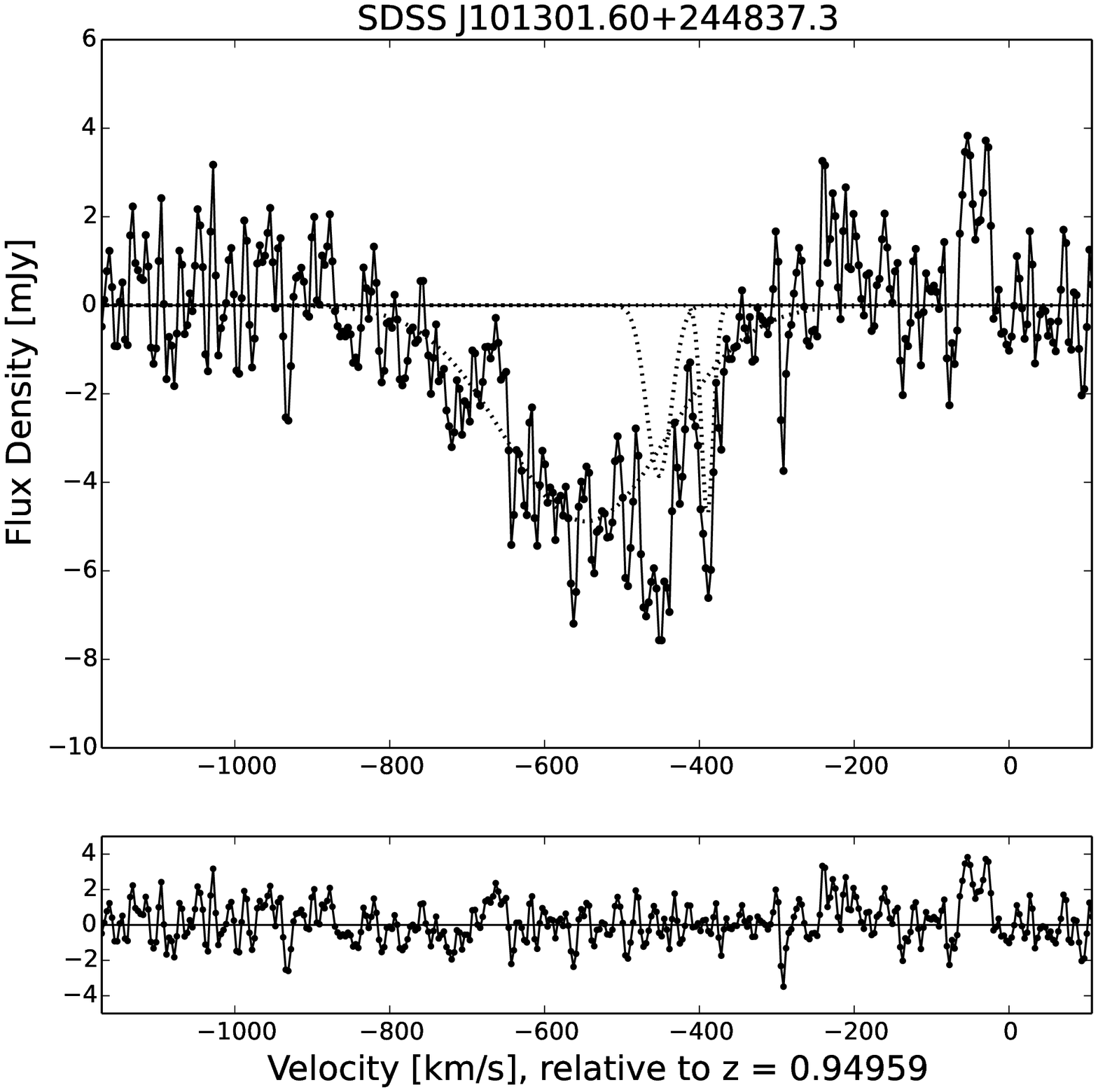} \\
\caption[]{The uGMRT detection of associated \hi absorption towards 1013+24, at $z = 0.94959$. Top panel shows the Gaussian fits 
to the absorption profile, and the panel at bottom shows the residuals after the fit has been subtracted from the spectrum.}\label{det3}
\end{figure}

\begin{figure}
\begin{tabular}{c}
\includegraphics[height=10cm, width=9cm]{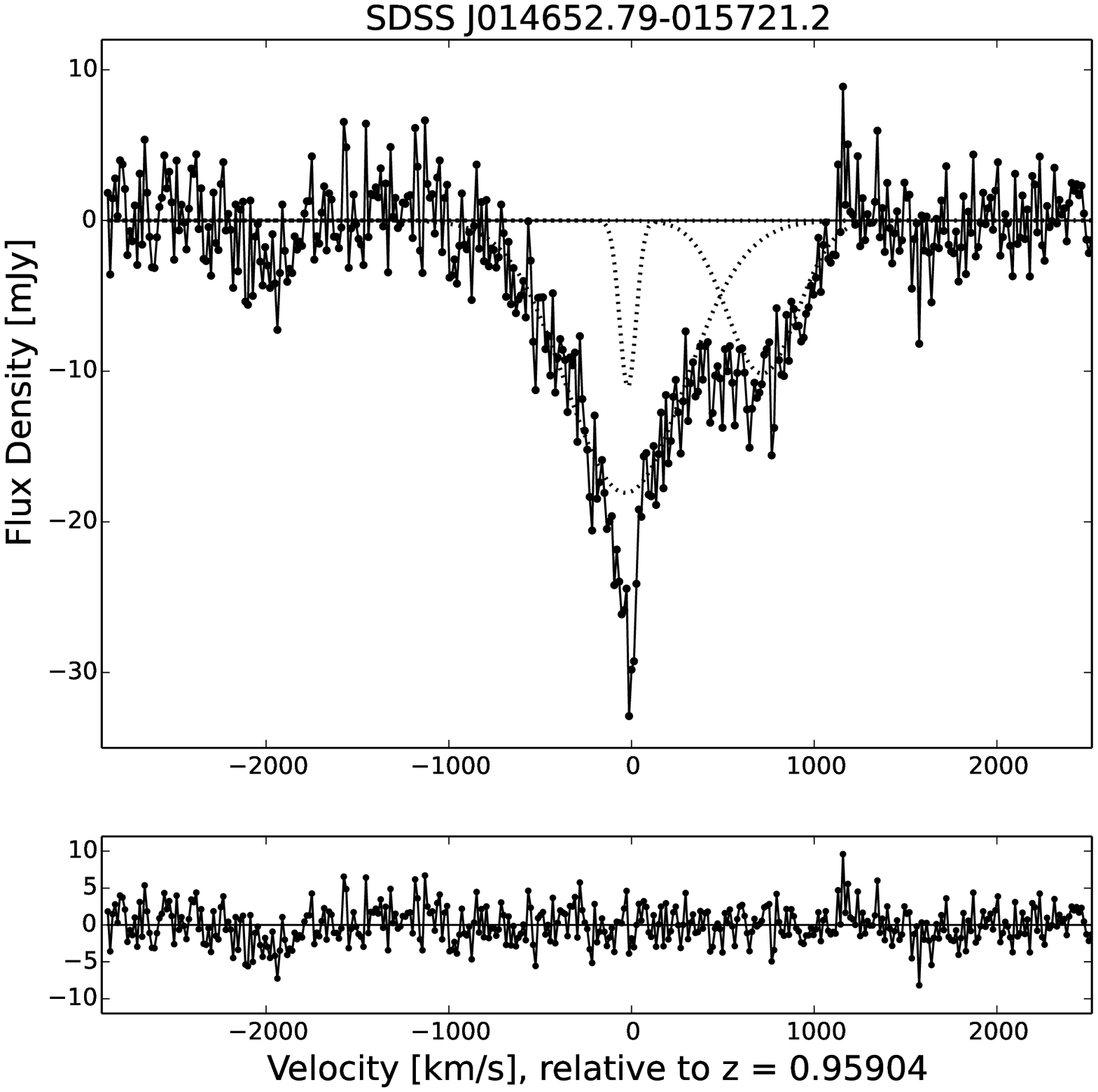} \\ 
\end{tabular}
\caption[]{The uGMRT detection of associated \hi absorption towards 0146-01, at $z = 0.95904$. Top panel shows the Gaussian fits 
to the absorption profile, and the panel at bottom shows the residuals after the fit has been subtracted from the spectrum. }\label{det4}
\end{figure}

\section{Sample selection}\label{selection}

The sample is based on the Faint Images of the Radio Sky at Twenty-Centimeters \citep[FIRST,][]{becker1995} catalogue.
Initially, the sources that are brighter than 100~mJy at 1.4 GHz were selected, since these 
would require shorter integration times (a few hours) to achieve reasonable \hi optical depth sensitivity. Next, the sky
coordinates of the sources were cross-matched with those in Sloan Digital Sky Survey (SDSS) Data Release 13 \citep[][]{albareti2017}, and based on the optical 
redshifts 590 sources were identified at $0.7 < z < 1.6$. These sources were observable with the Band-4 of uGMRT, to search for associated 
\hi absorption. Out of the 590 sources, 60 objects are identified as galaxies (all at $z < 1$), while the remaining 530 are identified 
as quasars, in the SDSS database. For the pilot survey presented here, a sample of 11 radio-bright galaxies was selected, all the sources in the sample are at $0.7<z<1.0$. 

Lately, the Wide-field Infrared Survey Explorer (WISE; \citealp[][]{cutri2013a, cutri2013} ) colours have been used by various authors to relate the radio source and/or \HI~ properties with those of the host galaxy \citep[e.g.][]{chandola2017, maccagni2017, moss2017}. The emission lines from dust and Polycyclic-Aromatic Hydrocarbons (PAHs), that may trace star formation activity in a galaxy, are known to peak at 11.3~$\mu$m \citep[e.g.,][]{nikutta2014}.
The W3(12~$\mu$m) band of WISE is hence known to be sensitive to the presence of dust and PAHs. Galaxies that are rich in dust have a enhanced luminosity at 12~$\mu$m, thus having a higher W2-W3 colour; starburst galaxies that are rich in PAHs and dust typically have W2-W3 > 3.4 mag \citep[e.g.,][]{rosario2013}.
Hence for the current sample, the sky coordinates of each source were cross-matched with Wide-field Infrared Survey Explorer (WISE; \citealp[][]{cutri2013a, cutri2013} ) all-sky survey catalogue to extract the WISE magnitudes, using the Vizier online catalogue access tool \citep[][]{ochsenbein2000}. The redshifts, 1.4 GHz flux densities and WISE magnitudes of the sample are presented in Table~\ref{17sam}. In the current sample, 0146-01 and 0910+24 have the highest W2-W3 colour, with 3.60 mag and 3.96 mag respectively, while the median W2-W3 colour of the sample is 2.89 mag.

\section{Observations and Data reduction}\label{observations}
 The observations were carried out during 15th May to 17th June, 2017 (proposal ID: $32\_ 101$). The uGMRT's 550-850 MHz receiver was used to observe the sample. The newly released GMRT Wideband Backend (GWB), with 200 MHz bandwidth covering 650-850 MHz, split into 8 K channels, was used.
This yielded a spectral resolution of $\sim 10$~km s$^{-1}$ at the centre of the band, sufficient to detect and resolve \hi absorption lines that typically have line-widths in the range of a few tens to a few hundreds of km s$^{-1}$. In three cases (1048+35, 1013+24 and 0146-01) weak, tentative  absorption lines were detected. These were followed up with narrow band observations using the GMRT Software Backend (GSB), with 4 MHz bandwidth for 1048+35 and 1013+24, and 16 MHz bandwidth for 0146-01. The bandwidths were split into 512 channels each, yielding a finer velocity resolution.
The final velocity resolution was $\rm \sim 20~km~s^{-1}$ for the non-detections, after Hanning smoothing and re-sampling (see discussion below), and $\rm 13.5~km~s^{-1}$, $\rm 3.4~km~s^{-1}$, $\rm 3.2~km~s^{-1}$ and $\rm 9.4~km~s^{-1}$,
for 0146-01, 1013+24, 1048+35 and 1648+22, respectively.

The data were recorded by following the usual pattern of observing a standard flux calibrator; i.e. 3C48, 3C147 or 3C286 every few hours. Long observing scans on the target source were bracketed by short observing scans on nearby phase
calibrator. Nearby compact sources, within $\approx 10^{\circ}-15^{\circ}$ of the target source, chosen from the VLA calibrator manual, were used for phase calibration. The flux calibrators were used to calibrate the system passband. Typically, the scans on the flux calibrators were of $\approx 10$ minutes, while those on phase calibrators were $\approx 7$ minutes.

The data were analysed using the Astronomical Image Processing System (AIPS, \citealp[][]{greisen2003}). The standard procedures of data editing, gain and bandpass calibration, self-calibration and imaging were followed. A more detailed description of the data reduction steps can be found in \citet[][]{aditya2016}, but note that the paper does not include the galaxies that are considered here. For the results presented here, only 17 MHz of the band, centered at the redshifted \hi frequency, was used. The procedure yielded a continuum image at the observing frequency, and a continuum subtracted \hi absorption spectrum extracted at the location of the target radio source. Except for four sources with detections all the spectra were Hanning-smoothed and re-sampled. The final spectrum was then obtained by fitting and subtracting a second order polynomial, if required, from the line free channels.

\begin{figure*}
\caption[]{The uGMRT \hi spectra for the 7 sources with non-detections. The continuum subtracted flux density (in mJy) is plotted against the heliocentric velocity (in km/s), relative to the source redshift. }\label{non_detections}
\begin{tabular}{ccc}

\includegraphics[scale=0.28]{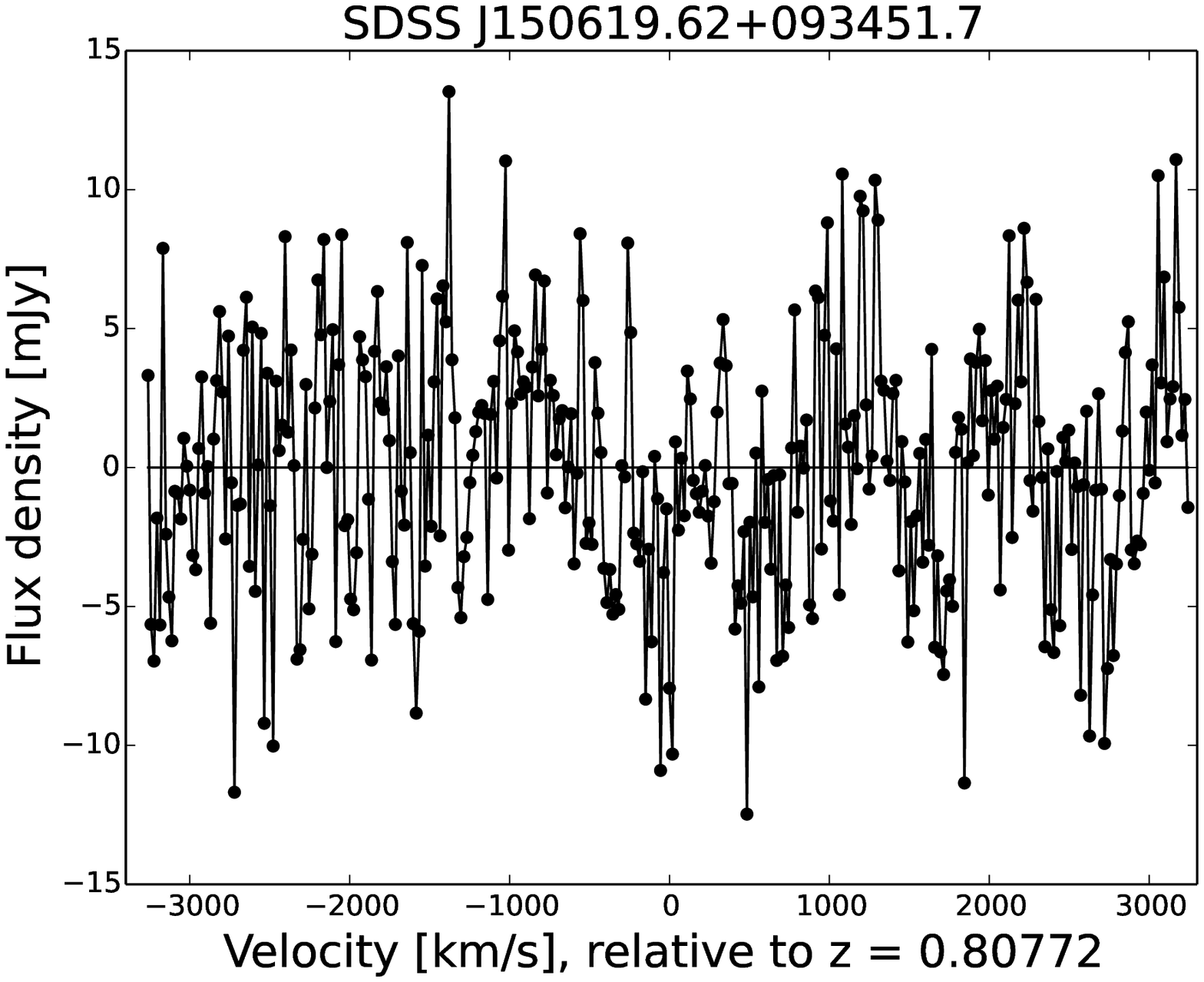} &
\includegraphics[scale=0.28]{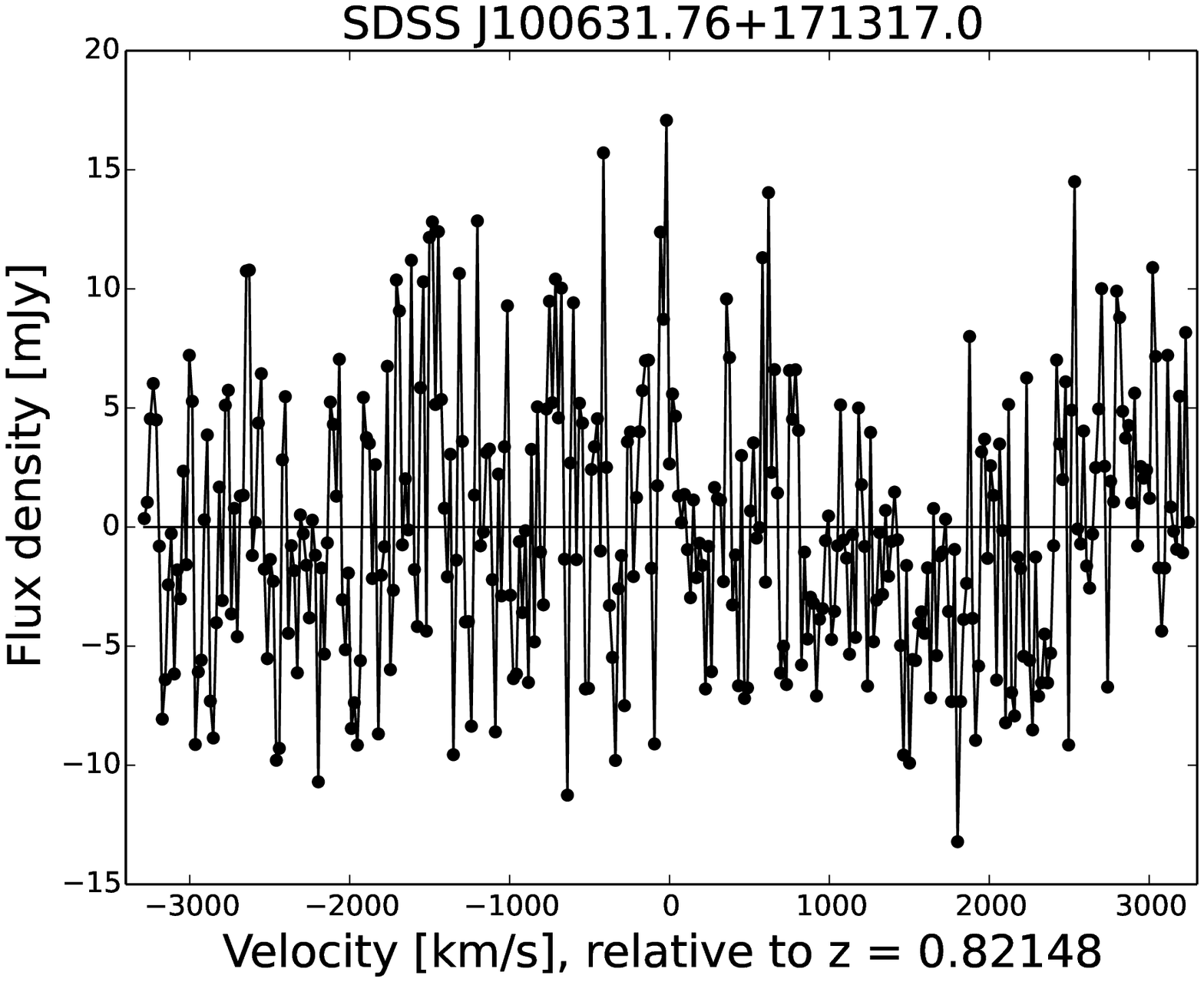} &
\includegraphics[scale=0.28]{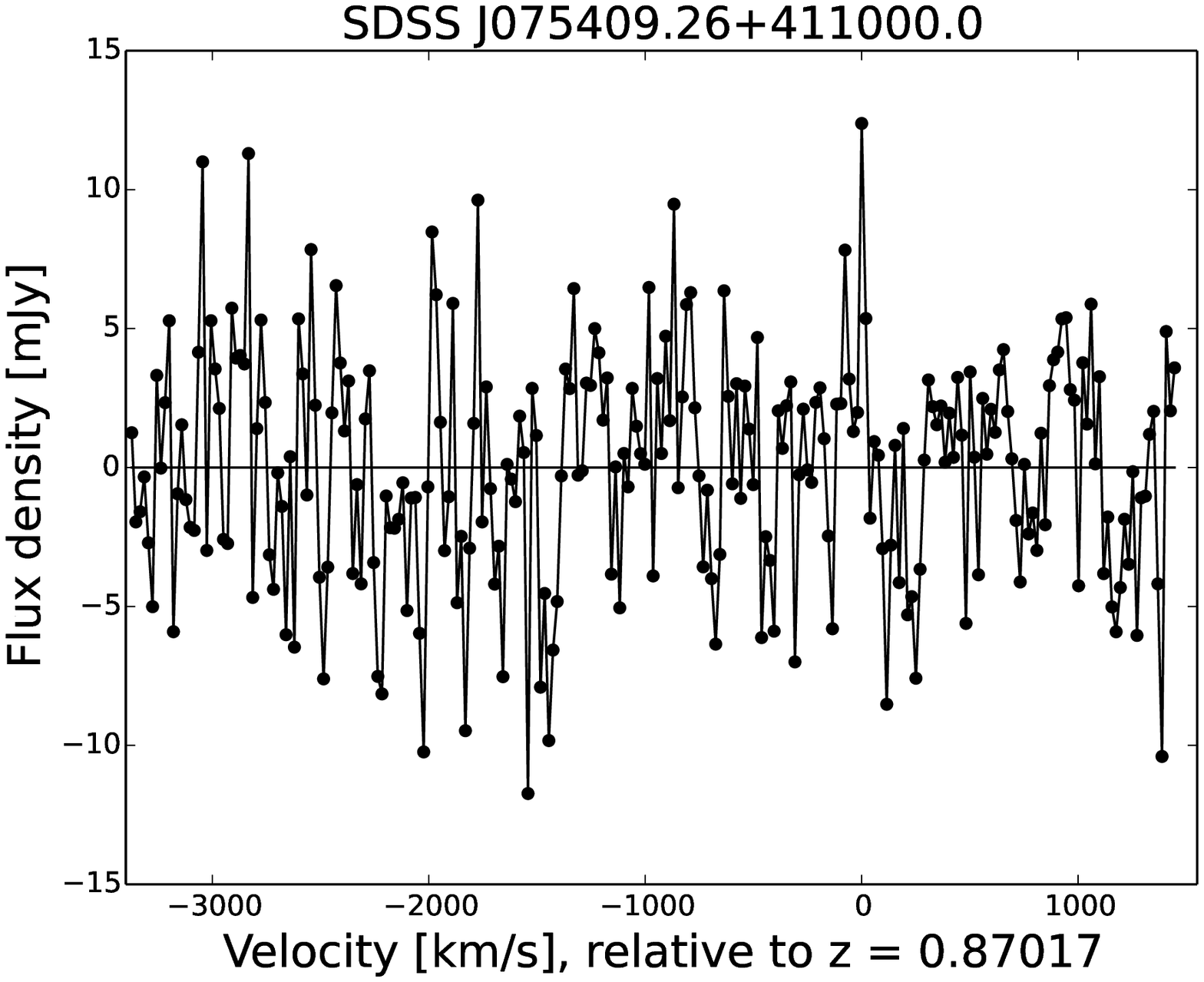} \\

\includegraphics[scale=0.28]{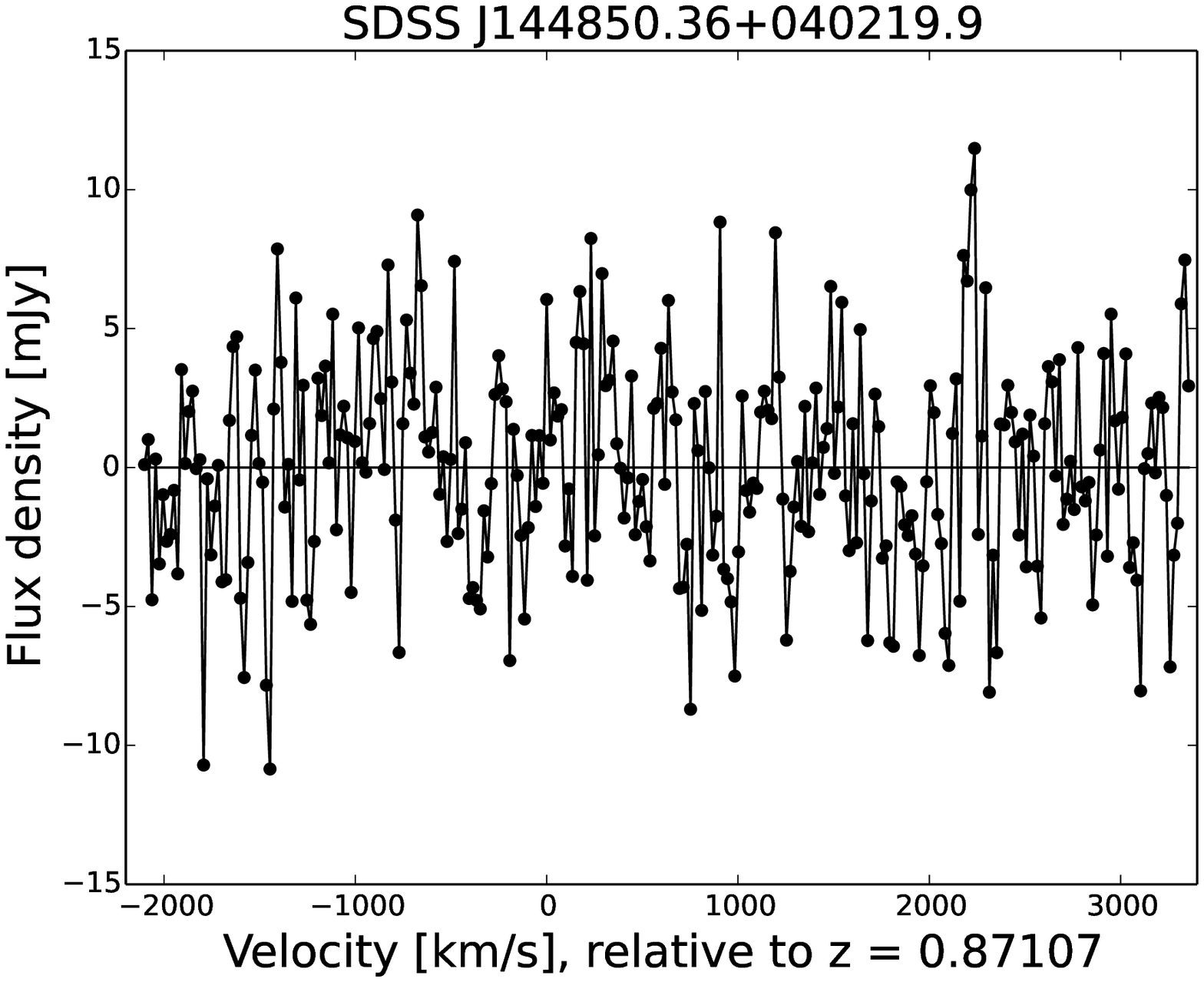} &
\includegraphics[scale=0.28]{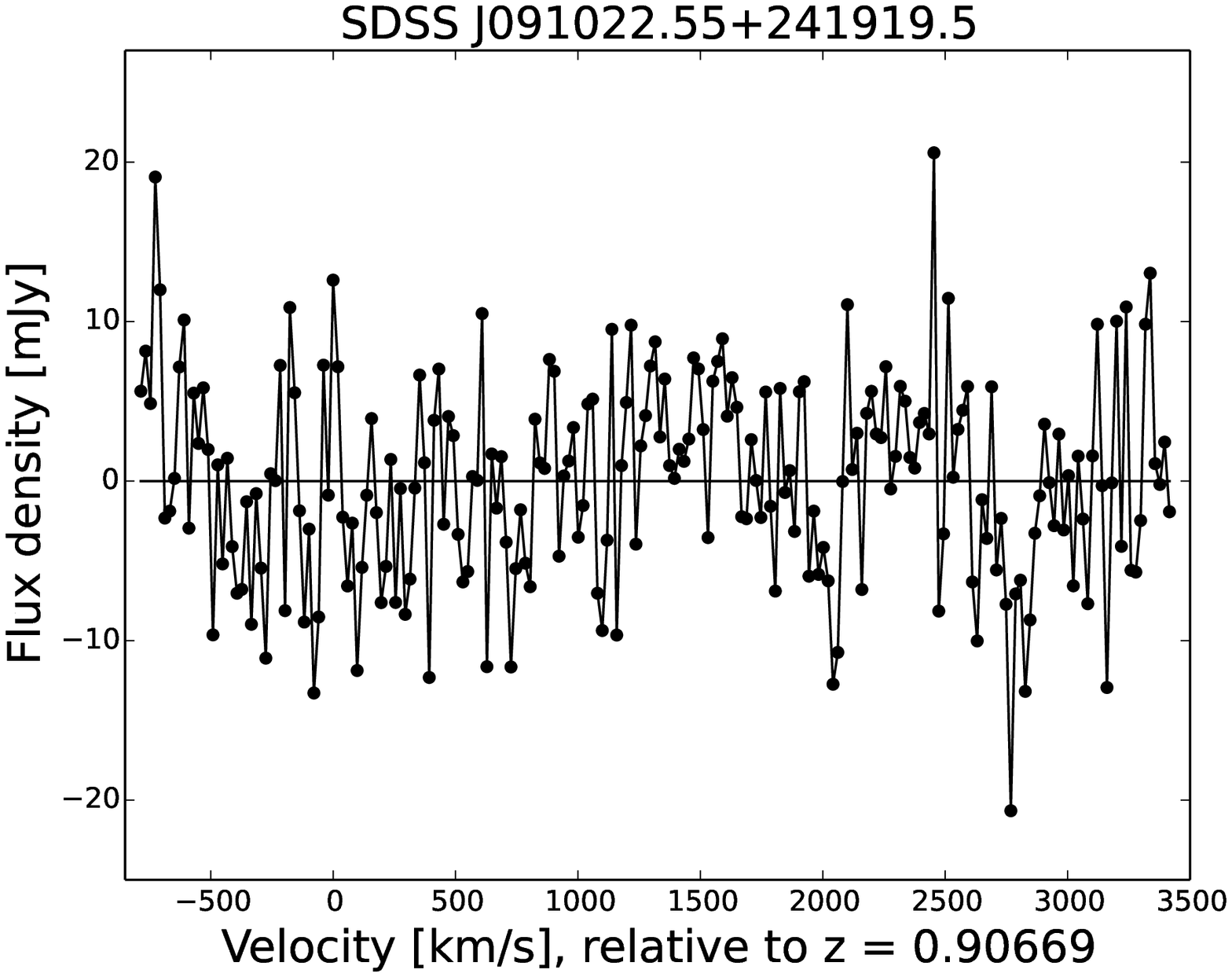} &
\includegraphics[scale=0.28]{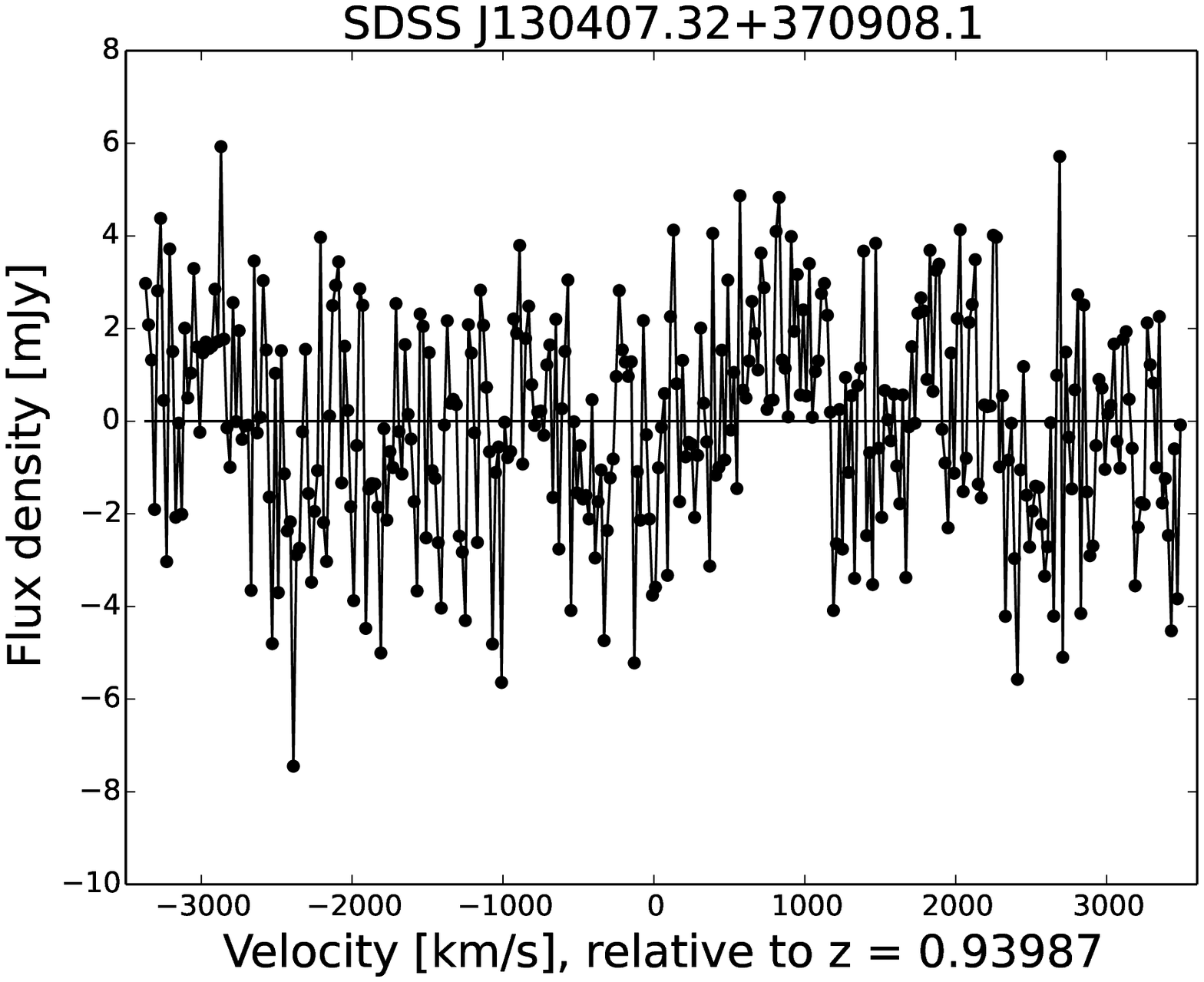} \\

\includegraphics[scale=0.28]{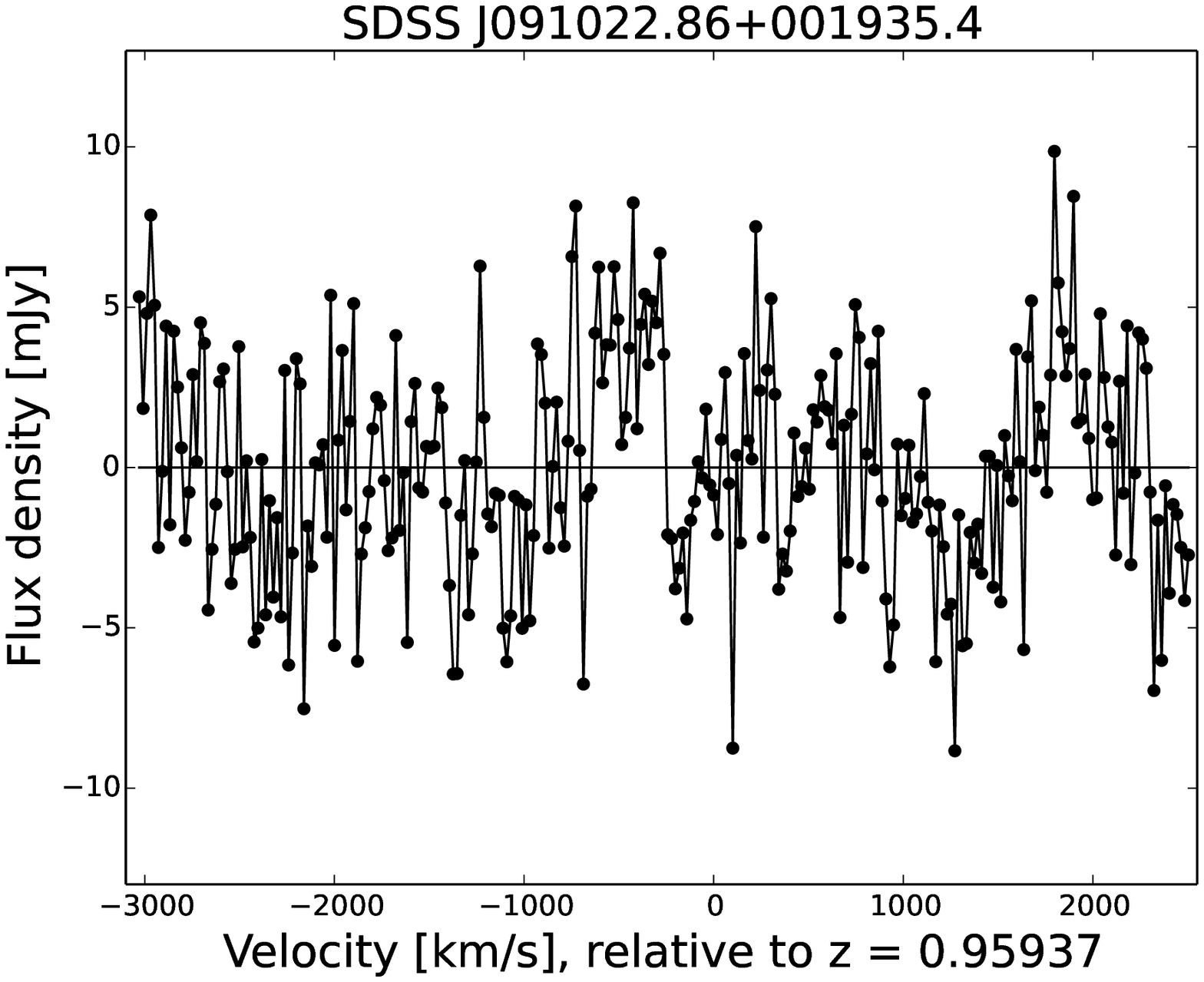} &

&
\\

\end{tabular}
\end{figure*}

\section{Results}\label{results}
  
The Band-4 uGMRT images have a spatial resolution of $\approx 6\arcsec~\times~6\arcsec$, all the radio sources were unresolved at this resolution.
A single Gaussian component was hence fitted to the target source in the image plane, and the peak flux densities (listed in Table~\ref{17gal}) were 
measured, using the task {\small{\textbf{JMFIT}}}. Here it should be noted that although {\small{\textbf{JMFIT}}} measurement errors are $< 1$~mJy in all cases, the errors on the flux densities are dominated by systematic effects ($\approx 10$\%), for GMRT at these frequencies, based on earlier observations. 
Four new detections of associated \hi absorption were obtained; towards 1648+22 at $z = 0.82266$, towards 1048+35 at $z = 0.84644$, towards 1013+24 at $z = 0.94959$ and towards 0146-01 at $z = 0.95904$. In the case of 1648+22 the absorption line was detected 
with a high Signal-to-Noise Ratio (SNR) in I, RR and LL stokes, with similar line strength in all cases. In the cases of 1048+35, 0146-01
 and 1013+24, the reality of the absorption feature was confirmed using a second uGMRT observing run, at a different epoch. The absorption feature has similar line strength in RR and LL stokes, and a high SNR, in all cases.
The spectra of the four detections are plotted in the Figures~\ref{det1},~\ref{det2},~\ref{det3}, and \ref{det4}. For the three sources 1048+35, 1013+24 and 0146-01 that have been re-observed, the spectra correspond to the second observing run that has a higher sensitivity. The spectra of the remaining 7 sources are consistent with noise, with no evidence of any significant absorption; the median RMS noise of the sample is 4.12 mJy.
The RMS noise on the final spectra are obtained on the line-free, RFI-free spectrum. The spectra of sources with non-detections are plotted in Figure~\ref{non_detections}. The panels in the figure show flux density (in mJy) plotted against velocity (in km s$^{-1}$) relative to the source redshift. 
Further, the integrated [O {\sc ii}] emission line luminosities (L$\rm _{OII}$) for all the 11 radio sources were estimated, using their optical spectra taken from SDSS \citep[][]{abolfathi2018}. For this, a spectrum with a width of 
$\approx 400$~\AA~around the [O {\sc ii}] emission line was extracted from the SDSS spectrum for each target source. A linear polynomial was fitted to the line-free spectrum and was subtracted. Then the integrated flux of the line was obtained by adding the fluxes on the emission line. Using the SDSS redshift and the integrated [O {\sc ii}] emission flux, line luminosity was estimated for each source. Note that here no correction was applied for dust reddening, which means that the quoted luminosities must be treated as lower limits.
 Table~\ref{17gal} summarizes the results from the GMRT observations, and lists the rest-frame 1.4 GHz radio luminosities and the [O {\sc ii}] line luminosities of the sources.

In the following, the details of the four new detections are discussed.
\newline
\newline
\textbf{1. 1648+22, at $z = 0.82266$ } 
\newline
\newline
The [O {\sc ii}], [O {\sc iii}] and H$\beta$ emission lines are clearly detected in the SDSS optical spectrum.
The redshift estimate for each source in the current sample was made using template fitting using a least-squares minimization algorithm \citep[][]{bolton2012, abolfathi2018}. The detection of a large number of
lines allows an accurate estimate of the AGN redshift. The estimated redshift of the source is $0.82266 \pm 0.00012$. The \hi absorption feature has a full width at nulls of $\approx 330$~km/s (see Figure~\ref{det1}). The integrated \hi optical depth of the absorption is $16.06 \pm 0.74$~km s$^{-1}$, implying a \HI~column density of $(29.28 \pm 1.35)$$\rm \times 10^{18}\times(T_{s}/cf)$ cm$^{-2}$. This is the highest \HI~column density detected in the sample. A single Gaussian profile has been fitted to the absorption feature, as shown in the Figure~\ref{det1}. The number of Gaussians fitted here, and for the remaining detections, was the minimum needed to reduce the post-fit residual to noise. A chi-square test was used to determine the `goodness' of the fit; the reduced $\chi^{2}$ value is 1.02.
The single Gaussian has a Full Width at Half Maximum (FWHM) of $194$ km s$^{-1}$, that is marginally wide compared to \hi absorbers at low redshifts \citep[e.g.,][]{maccagni2017}. The centroid of the Gaussian is redshifted relative to the AGN redshift by 107 km s$^{-1}$ (see Table~\ref{8det}).

The spectrum of the source is extremely flat at low radio frequencies; the spectral index ($\alpha$; $S_{\nu} \propto \nu^{\alpha}$) between 1.4 GHz 
(flux density obtained from \citet[][]{becker1995}, FIRST radio catalogue) and 779 MHz (from current observations) is -0.02. 
This suggests that the radio source is either compact or core-dominated since such a flat spectral shape mainly occurs due to synchrotron
self-absorption occurring in a relatively compact and dense emitting region.
Indeed, the radio source is resolved only to a small degree in the milli-arcsec scale VLBI image at 4.8 GHz, by \citet[][]{helmbolt2007}.
The source has a bright core with a flux density of $\approx 346$~mJy at 4.8 GHz \citep[][]{helmbolt2007}, along with a faint one-sided radio lobe extending up to $\approx 69$~pc towards west. A bulk of the \hi absorption is likely arising against the bright radio core that has relatively higher flux density. This is because the low flux density of the extended lobe would imply a high \hi opacity to produce the observed \hi absorption. A strong absorption against the faint lobe would mean that there is an abrupt increase in the \HI~opacity along the lobe. However, a contribution to the absorption from gas in front of the diffuse extended emission cannot be excluded in case of a strong absorption against the core. The simplicity of the absorption profile and near coincidence with the optical redshift (i.e. peak of absorption is within 200 km s$^{-1}$) suggests that it possibly originates from the gas in the circumnuclear disk.
\newline
\newline
\textbf{2. 1048+35, at $z = 0.84644$ } 
\newline  
\newline
Strong [O {\sc ii}], H$\beta$, [Ne {\sc iii}] and [C {\sc ii}] emission lines are clearly detected in the optical spectrum of the source, and the redshift has been estimated to be $z = 0.84644 \pm 0.00006$ \citep[][]{abolfathi2018}. The spectrum at low radio frequencies is relatively steep, with $\alpha = -1.20$ between 1.4 GHz and 769 MHz. 
High angular resolution VLBI studies are currently not available for this source in the literature. The \HI~absorption spectrum from the second observation, that has a finer velocity resolution of 3.2 km s$^{-1}$, is displayed in the left panel of Figure~\ref{det2}. The spectrum shows a `two-horned' absorption profile, with two absorption peaks nearly centred around the AGN redshift. Along with this, a shallow blueshifted absorption `tail' extending up to -300 km s$^{-1}$ can also be seen.
Five Gaussian components were fitted to the absorption profile, as shown in Figure~\ref{det2}; the reduced $\chi^{2}$ value is 0.90. Three components are relatively narrow, with FWHM $< 40$ km s$^{-1}$, and two components are relatively broad, one having a FWHM of 146 km s$^{-1}$, and the other having 311 km s$^{-1}$ (see Table~\ref{8det}). 
The two-horned absorption peaks have low velocity offsets ($\rm < 200~km~s^{-1}$) relative to the AGN.
Such absorption features mostly arise from \HI~clouds that are possibly rotating in circumnuclear disks. The broad blueshifted feature, with a velocity offset of $\rm \approx 300~km~s^{-1}$ at the null, is possibly the resultant of AGN jet-gas interactions or a gas-rich merger, as discussed above.   
\newline
\newline
\textbf{3. 1013+24, at $z = 0.94959$ } 
\newline
\newline
The radio source is unresolved in the uGMRT image at 728 MHz. High-resolution VLBI images for the source are currently not available in the literature. 
The spectrum at low radio frequencies is marginally steep, with a spectral index of -0.85 between 1.4 GHz and 728 MHz.
The SDSS optical image looks redder, with a slightly diffused structure \citep[][]{abolfathi2018}. Prominent [O {\sc ii}], [O {\sc iii}] and [Ne {\sc iii}] emission lines are clearly detected in the optical spectrum. The estimated redshift of the source is $z = 0.94959 \pm 0.00008$ \citep[][]{abolfathi2018}. Three Gaussian profiles were fitted to the absorption feature; one of the components is broad with a FWHM of 247 km s$^{-1}$, and the other two  are relatively narrow, with FWHM < 40 km s$^{-1}$ (see Table~\ref{8det}). The reduced $\chi^{2}$ value is 0.87. Interestingly, the peak of the \HI~absorption profile is significantly blueshifted from the AGN optical redshift by $\approx 450$~km s$^{-1}$ ($z = 0.94959 \pm 0.00008$). Strong blueshifted absorption features often show signatures of weaker extended absorption wings, indicating diffused gaseous outflows resulting from the kinematic interaction of AGN jet with gas. The diffused outflows tend to have relatively lower velocities compared to the bulk outflow. However, in this case, the absorption profile looks nearly symmetric around the peak, with no detection of any weaker profile near the AGN systemic velocity. The present GMRT \HI~absorption spectrum is possibly not sensitive enough to detect any weak absorption wing near the systemic velocity. Or in an alternative case, the symmetric absorption could be representing a kinematically `un-disturbed' atomic cloud with no diffused gaseous structure, intercepting the line of sight towards the radio source at a lower redshift.  
\newline
\newline
\textbf{4. 0146-01, at $z = 0.95904$ } 
\newline
\newline
A slew of both narrow and broad emission lines that include [O {\sc iii}], [O {\sc ii}], H$\gamma$, H$\beta$, [Ne {\sc iii}], Mg, [C {\sc ii}] etc., are detected in the optical spectrum. The redshift has been estimated to be $z = 0.95904 \pm 0.00005$ \citep[][]{abolfathi2018}. High spatial resolution VLBI studies are not available for this source in the literature.  The spectrum at low radio frequencies is relatively steep, with $\alpha = -1.12$ (estimated between 1.4 GHz and 725 MHz), suggesting the possibility of extended radio structure. However, the source is unresolved in the uGMRT image with $\approx 6 \arcsec \times 6 \arcsec$ resolution. The \hi absorption is extremely wide, with a width of $\approx 1800$~km s$^{-1}$ at the absorption nulls (see Figure~\ref{det4}). The total velocity integrated optical depth of the absorption profile is $11.46 \pm 0.22$ km s$^{-1}$, implying a high \HI~column density of $(20.89 \pm 0.04)$$\rm \times 10^{18}\times(T_{s}/cf)$ cm$^{-2}$. 

The absorption has a complex profile with three Gaussian components, with two of them having broad FWHM of over 400 km s$^{-1}$; the reduced $\chi^{2}$ value is 0.91. The centroids of two components are marginally blueshifted relative to the AGN, while the third component is redshifted by 715 km s$^{-1}$ (see Table~\ref{8det}), suggesting that a significant fraction of the gas mass is possibly flowing towards the AGN. Earlier, a redshifted absorption with a similarly large velocity of 636 km s$^{-1}$ relative to the AGN, was detected towards PKS 0428+20 at $z = 0.219$, by \citet[][]{vermeulen2003}. Recently, \citet[][]{dutta2018} have also reported large velocity inflows in merging systems at $z \lesssim 0.2$. Such large gas inflows could be explained in scenarios of galaxy mergers, wherein gas components get channelled to the central regions due to removal of angular momentum \citep[e.g.,][]{jogee2006, dimatteo2007}. Circumnuclear starbursts are often seen to be associated with such a process; a high WISE colour (W2-W3) of 3.60 mag is also indicative of this. However, further high-resolution VLBI studies at the redshifted \hi frequency are needed to understand the origin of this component.
  
Further, the absorption null extends up to $\approx 700$ km s$^{-1}$ on the blue side, implying that a significant fraction of the gas mass that has negative velocities, is also outflowing relative to the AGN.
Curiously, the detection towards TXS 1245-197 at $z = 1.275$, reported recently by \citet[][]{aditya2018a}, has a similar looking blueshifted absorption wing, that extends up to $\approx -800$~km s$^{-1}$. The absorption peak is also coincident with the AGN redshift, as in the current case. It is argued in the case of TXS 1245-197 that the cold gas outflow is possibly driven by the radio jet,
which could be the case even in the current system since such extremely fast outflows are expected to be the resultants of jet interactions \citep[e.g.,][]{morganti2013, maccagni2017}. Earlier, there have been many such detections with outflows reported in the literature, and importantly fast cold gas outflows (with velocity $\gtrsim 500$~km s$^{-1}$) are more commonly seen to be originating from the central kpc region, around the AGN \citep[e.g.,][]{morganti2018}. Possibly, the impact of the radio jets on the ambient medium could be maximum in these regions.
  
Next, by assuming that the outflow of neutral gas in the current system is driven by a mass-conserving free wind, I estimate the mass outflow rate ($\dot{\rm M}$) of \HI~gas using the following expression \citep[][]{morganti2005,heckman2002},

\begin{equation} \label{eu_eqn}
\dot{\rm M}  = 30.\Big[ \frac {\Omega} {4 \pi} \Big].\Big[ \frac { {\rm r}_{\star} } { 1~{\rm kpc}} \Big] .\Big[ \frac {\rm N_{HI}} { 10^{21}~ {\rm cm}^{-2}} \Big] .\Big[ \frac {\rm V_{out}} { 300~{\rm km}~{\rm s}^{-1} } \Big]. {\rm M}_{\odot}~{\rm yr}^{-1}            
\end{equation}

Here, $\rm V_{out}$ is the velocity of the outflow with which the gas flows into a solid angle $\Omega$, from a minimum radius ${\rm r}_{\star}$. Following \citet[][]{morganti2005}, I assumed a solid angle ($\Omega$) of $\pi$ radians, and the minimum radius ${\rm r}_{\star}$ to be 1~kpc. The outflow velocity is assumed to be equal to half the full width at zero intensity, relative to the AGN systemic velocity (i.e. half the velocity width from the outer edge of the blueshifted \hi absorption to the AGN systemic velocity); this corresponds to $\approx 350$~km s$^{-1}$. A spin temperature of 1000 K and a covering factor of unity are assumed in estimating the \HI~column density, $\rm N_{HI}$. I note here that these assumptions allow a direct comparison with the low-$z$ results of \citet[][]{morganti2005}. Also, a high spin temperature ($\rm \sim 1000~K$) is indeed expected for neutral gas that is located close to the AGN \citep[e.g.][]{maloney1996}. Further, \hi absorption studies of intervening damped Lyman $\alpha$ systems have pointed out that galaxies at high redshifts typically have higher spin temperatures, indicating larger fractions of warm neutral medium \citep[e.g.][]{kanekar2014}, which may well also be the case for AGN environments. The estimated mass outflow rate, $\dot{\rm M}$, is $\approx 78$~{\rm M}$_{\odot}$~yr$^{-1}$, which is the highest compared to the cold \HI~mass outflow rates estimated in both high and low redshift systems to date \citep[e.g.,][]{morganti2018}. However, I note that this estimate critically depends upon the assumptions of 
$\rm T_{s}$, ${\rm r}_{\star}$ and $\Omega$. The typically known outflow scales (${\rm r}_{\star}$) range from few$\times 10$ pc to $\gtrsim$ 1 kpc \citep[e.g.][]{morganti2005}. Hence, as such, the estimated mass outflow rate would represent an upper limit, considering high $\rm T_{s}$ and ${\rm r}_{\star}$ values, of 1000 K and 1~kpc respectively. Lower values, for example, a spin temperature of 100 K and ${\rm r}_{\star}$ of 100 pc, would yield a lower $\dot{\rm M}$.

\section{Discussion}

\begin{table*}
\footnotesize
\caption{Gaussian fit parameters for the four new detections.} 
\label{8det}
\begin{scriptsize}
\begin{tabular}{ccccc} 

\hline \\
Source &  $ z$$^{a}$ & $\int \tau dv$ & $V_{HI}$$^{b}$ & $\Delta V$$^{c}$  \\
       &        &    [km s$^{-1}$]       & [km s$^{-1}$] & [km s$^{-1}$]   \\

\hline

 1648+22 & $0.82266 \pm 0.00012 $& $16.06 \pm 0.74$& +107     & 194   \\ 

 1048+35 & $0.84644 \pm 0.00006$ & $4.77 \pm 0.18$ & +114     & 17    \\
                          &                       &                 & +75      & 38    \\
                          &                       &                 & +25      & 146   \\
                          &                       &                 & -40      & 22    \\
                          &                       &                 & -198     & 311   \\

 1013+24 & $0.94959 \pm 0.00008$ & $1.15 \pm 0.05$ & -547     & 248   \\
                          &                       &                 & -452     & 38    \\     
                          &                       &                 & -390     & 18    \\
 0146-01 & $0.95904 \pm 0.00005$ & $11.46 \pm 0.22$& -33      & 762   \\
                          &                       &                 & -20      & 105   \\
                          &                       &                 & +715     & 450   \\

\hline
\hline

\end{tabular}
\end{scriptsize}

\begin{tablenotes}
\item[]~Notes: 
\item[a]$^{a}$~ Reference for redshift:  \citet[][]{abolfathi2018}.
\item[b]$^{b}$~ Velocity offset of the absorption line centroid from the optical redshift (negative means blueshifted line).
\item[c]$^{c}$~ The line full width at half maximum.

\end{tablenotes}
\end{table*}

\subsection{ Excess [O {\sc ii}] emission in associated absorption systems ?}
There have been many efforts in the past that have used the optical emission lines to understand the 
AGN and the host galaxy properties \citep[e.g.][]{baldwin1981, tadhunter1998}. While broad and high ionization lines tend to arise from the immediate vicinity of the central massive blackhole, low ionization lines arise further out \citep[e.g.][]{richardson2014}, tracing the host-galaxy dynamics, chemical composition and gas distribution. It is known that low ionization gases trace the \HI~distribution in the galactic disks \citep[e.g.][]{lin2001}. Indeed, Mg {\sc ii} is known 
as a good tracer of high column density \HI~gas in damped Lyman $\alpha$ absorbers \citep[e.g.][]{bergeron1986, peroux2004}. [O {\sc ii}] $\lambda~3727$ is widely used to trace star formation in galaxy surveys, particularly for redshifts $z \gtrsim 0.4$ \citep[e.g.][]{lilly1996, hippelein2003}. In optically selected quasars, [O {\sc ii}] emission in excess of the base level that is expected from non-stellar photoionization indicates star formation accompanying quasar activity \citep[e.g.][]{ho2005}.
These lines will hence allow a better understanding of the distribution and kinematical properties of the atomic gas.
Recently, \citet[][]{shen2012} and \citet[][]{khare2014} studied the properties of [O {\sc ii}] $\lambda~3727$ emission from a large sample of associated Mg {\sc ii} absorbers, in order to understand the origin of associated Mg {\sc ii} absorption systems relative to the AGN. \citet[][]{khare2014} find a high excess [O {\sc ii}] emission in the composite quasar
spectrum (constructed in the quasar rest frame) for the sample of quasars that show outflows, i.e. the systems for which the Mg {\sc ii} absorption is blueshifted relative to the AGN. Further, the authors find that the presence of an associated Mg {\sc ii} absorption enhances [O {\sc ii}] emission in the composite quasar spectrum. However the excess emission flux in the [O {\sc ii}] line does not depend on the strength of the Mg {\sc ii} absorption line \citep[][]{khare2014}. The excess [O {\sc ii}] line flux could be originating either from the host galaxy, or the parent AGN; in the former case, the excess [O {\sc ii}] emission could indicate higher star formation in the host galaxy. The association of high [O {\sc ii}] line emission with the occurrence of outflows suggests either strong stellar outbursts in the host galaxy, or strong AGN jet-gas interactions.

\begin{figure}

\includegraphics[scale=0.45]{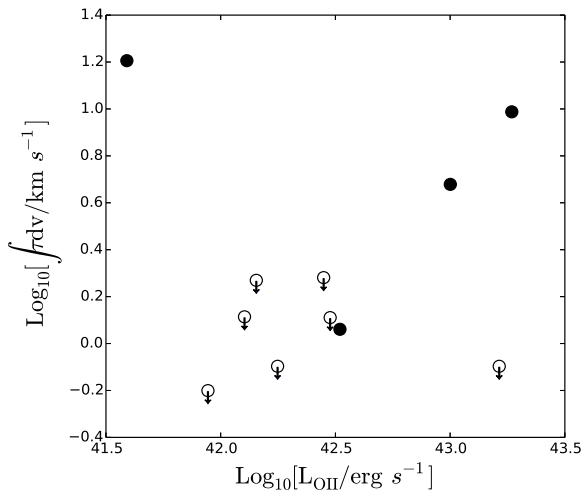} \\
\caption[]{ The integrated optical depths plotted as a function of [O {\sc ii}] luminosity, ${\rm L}_{\rm OII}$. Hollow circles represent upper limits  while solid circles represent detections of \hi absorption.  } \label{o2_tau}

\end{figure}

To test this scenario in the case of associated cold \HI~absorption systems, that are the focus of this paper, the integrated \hi optical depths are plotted with respect to the L$\rm _{OII}$ in Figure~\ref{o2_tau}. Interestingly, all the three \hi detections with blueshifted features (0146-01, 1048+35 and 1013+24) have relatively high L$\rm _{OII}$ compared to a bulk of the remaining sample. 0146-01 with a fast cold outflow ($\approx 700$ km s$^{-1}$) has the highest L$\rm _{OII}$ in the sample, while 1048+35 and 1013+24 have the third and fourth highest L$\rm _{OII}$ respectively. These results are consistent with the findings of \citet[][]{khare2014}. 

In the case of 0910+24, the source has the second highest L$\rm _{OII}$ and also the highest WISE colour (W2-W3 = 3.96 mag)
in the sample, both suggesting that the host galaxy is likely harbouring extreme star-burst activity, and possibly rich in PAHs and dust. It is surprising that \hi absorption is not detected towards the radio source, particularly in view of the fact that the source 0146-01, with \hi detection, has a similarly high W2-W3 colour and L$\rm _{OII}$. It is likely that the neutral gas is at a higher spin temperature, or it is not obscuring our line of sight towards the radio source in the case of 0910+24. Further, the detection towards 1648+22, although has the highest \HI~integrated optical depth in the sample, has the least L$\rm _{OII}$. It should be noted here that there is no signature of a strong \HI~outflow in the uGMRT spectrum of 1648+22. The peak of the \hi absorption is redshifted relative to the AGN by 193.5 km s$^{-1}$, and the absorption has a FWHM of $\approx 107$~km s$^{-1}$. The absorption extends only up to -75 km s$^{-1}$ on the blue side. Further, since the excess emission flux in the [O {\sc ii}] line does not seem to depend on the strength of the absorption line in the sample of Mg {\sc ii} absorbers studied by \citet[][]{khare2014}, a high \hi optical depth need not necessarily yield a high L$\rm _{OII}$. The source has W2-W3 colour of 3.05 mag, that is marginally higher than the median of the sample, suggesting the presence of dust. The weaker [O {\sc ii}] emission, in this case, could be purely of stellar origin. Finally, I emphasize that the current sample size is small, and a further study using a larger, statistically significant sample would reveal a possible connection between the occurence of \HI~outflows and [O {\sc ii}] emission.

\subsection{Distribution and kinematics of \HI}

As discussed above, currently the intermediate redshift range ($0.7 < z < 1.0$) is barely explored in terms of associated \hi absorption searches.
With 4 detections \citep[][]{vermeulen2003, salter2010, yan2016} and 10 upper limits, the number of such searches is just a handful. Current observations have added 4 new detections and 7 upper limits, doubling the number of detections at these redshifts, and also enhancing the sample size.  
As described in Section~\ref{selection}, the sample is purely flux-selected, based on SDSS and FIRST catalogues, similar in construction to 
\citet[][]{gereb2015} and \citet[][]{maccagni2017} samples at low redshifts, $z < 0.23$. However, here brighter sources with a higher median radio
power, ${\rm L_{1.4~GHz, med}} = 10^{27.06}~{\rm W}~{\rm Hz}^{-1}$ (see Table~\ref{17gal}), are targetted, whereas the sample by \citet[][]{maccagni2017} has sources with a median ${\rm L_{1.4~GHz}}$ of $10^{24}~{\rm W}~{\rm Hz}^{-1}$. The overall detection rate at low redshifts (at $z < 0.23$) has been observed to be $27\% \pm 5\%$, in a sample of 248 radio sources. \citet[][]{maccagni2017} also find that the detection rate is similar across the radio power, albeit with small number statistics in the bins, with the sample covering over three orders of magnitude in radio power. The current results are consistent with this, with a tentatively similar detection rate at a higher median radio power. However, since the current sample covers relatively small range of radio luminosities, $\rm 10^{26.66}~W~Hz^{-1}$ to $\rm 10^{27.67}~W~Hz^{-1}$ (see Table~\ref{17gal}), and also since the size of the sample is small, it would not be reasonable to separate the sample into bins and estimate the detection rate.
 
Next, the present observations are relatively shallow, with a median $3\sigma$ upper limit on the integrated optical depth of 2.04 km s$^{-1}$ for an assumed line width of 100 km s$^{-1}$, while it is 0.81 km s$^{-1}$ for the \citet[][]{maccagni2017} sample. A primary reason for the relatively shallow pilot search is that just 16 or fewer GMRT antennas were available during the observations in cycle 32 of GMRT. From cycle 33 onwards all the 30 antennas of GMRT will be fitted with UHF 550-850 MHz band receivers, and hence our future observations will probe deeper in terms of optical depth limits. The occurrence of four detections of associated \hi absorbers in a sample of 11 in the relatively shallow pilot survey is encouraging, in the sense that the detection rate at intermediate redshifts could be similar to that at low redshifts, $\approx 30\%$, on contrary to a significantly low detection rate observed in samples searched at $z > 1$ \citep[e.g.,][]{curran2013, aditya2016, aditya2018}.

Further, as discussed in Section~\ref{results} three detections show strong blueshifted features, suggesting \HI~outflows. 
The absorption null extends up to -300 km s$^{-1}$ in the case of 1048+35, -750 km s$^{-1}$ in 1013+24, and up to -700 km s$^{-1}$ in the case of 0146-01. Studies of low-$z$ populations by \citet[][]{vermeulen2003, gupta2006, gereb2015, maccagni2017} find that blueshifted absorptions more often occur towards powerful radio luminous AGNs, suggesting that outflowing gas motions in higher luminosity sources are caused due to interactions between the radio source and the surrounding gaseous medium. For comparison, in the low-$z$ sample of \citet[][]{maccagni2017}, the source with the largest blueshift of $284$~km s$^{-1}$ has a radio luminosity of ${\rm L}_{1.4~GHz} \approx 10^{26.15}$~W Hz$^{-1}$. In their sample, the absorbers with luminosities ${\rm L}_{1.4~GHz} > 10^{24.5}$~W Hz$^{-1}$ show blueshifts with $\gtrsim 200$ km s$^{-1}$. Hence, it is indeed expected to find a higher incidence rate of outflows in the current sample since relatively powerful radio sources, with 
median ${\rm L_{1.4~GHz}} $ of $ 10^{27.06}~{\rm W}~{\rm Hz}^{-1}$, were targetted. Particularly, all the three radio sources with blueshifted absorption features have ${\rm L_{1.4~GHz}}  > 10^{27}~{\rm W}~{\rm Hz}^{-1}$ (see Table~\ref{17gal}). These lie at the extreme high luminosity end of the distribution studied by \citet[][]{maccagni2017}, suggesting the powerful nature of the central engines in these objects. The \HI~outflows in these three cases are arguably caused due to radio jet-gas interactions.

\section{Summary}

A sample of 11 radio-bright galaxies at redshifts $0.7 < z < 1.0$ were observed using uGMRT to search for associated \hi absorption. 
The search has yielded four new detections and seven upper limits, increasing the number of known detections at intermediate redshifts to eight, 
and also enhancing the sample size of associated \hi absorption searches at these redshifts. The numbers indicate that the detection rate could be as high 
as that at low redshifts, $\approx 30\%$, on contrary to a low detection rate observed in samples at $z > 1$. 
Further, out of the four new detections, three show evidence for large blueshifted features, indicating cold gas outflows.
Highest cold gas mass outflow rate of $\approx 78$~{\rm M}$_{\odot}$~yr$^{-1}$ is estimated for the detection towards 0146-01,
at $z = 0.95904$, assuming a spin temperature of 1000 K. Further, the three absorbers with \HI~outflows also tentatively show an excess [O {\sc ii}] line luminosity than the bulk of the remaining sample, indicating that the hosts of these AGNs harbour different environments, either due to excess stellar outbursts or due to AGN jet-gas interactions. Studies of low-$z$ populations find that AGN associated \HI~outflows more commonly occur towards powerful central radio sources, arguing that the interactions between the radio source and the surrounding gas as the cause for outflowing gas
motions. The radio sources with outflows in the current sample are highly luminous, with L$_{\rm 1.4~GHz} > 10^{27} {\rm W}\, {\rm Hz}^{-1}$.
The extreme powerful nature of the central engines in these objects possibly explains the \HI~outflows in these sources.

\section*{ACKNOWLEDGEMENTS}
I thank R. Srianand and Neeraj Gupta for their invaluable comments and inputs, that helped in improving the paper significantly. I also thank the staff of the GMRT who have made these observations possible. The GMRT is run by the 
National Centre for Radio Astrophysics of the Tata Institute of Fundamental Research. This research has made use of the 
NASA/IPAC Extragalactic Database (NED) which is operated by the Jet Propulsion Laboratory, California Institute of Technology, under contract with the National Aeronautics and Space Administration. Further, I acknowledge support from the Indo-French Centre for the Promotion of Advanced Research under Project 5504-B (PIs: N. Gupta, P. Noterdaeme). I thank the anonymous referee for a very helpful report that significantly improved the quality of the paper.

\bibliographystyle{mnras}
\bibliography{ms}

\bsp	
\label{lastpage}
\end{document}